\documentclass[article,amsfonts, amssymb, amsmath, showkeys, nofootinbib, prx, superscriptaddress, preprintnumbers, twocolumn,longbibliography]{revtex4-2}
\usepackage{float}
\makeatletter
\let\newfloat\newfloat@ltx
\makeatother

\usepackage[english]{babel}
\usepackage[T1]{fontenc}
\usepackage[utf8]{inputenc}

\usepackage{graphics}
\usepackage{selinput}
\usepackage[normalem]{ulem}
\usepackage[shortlabels]{enumitem}
\usepackage{braket}

\usepackage{braket}
\usepackage{amsthm}
\usepackage{mathtools}
\usepackage{physics}
\usepackage{xcolor}
\usepackage{graphicx}
\usepackage[left=16mm,right=16mm,top=35mm,columnsep=15pt]{geometry} 
\usepackage{adjustbox}
\usepackage{placeins}

\usepackage{lipsum}
\usepackage{csquotes}
\usepackage{bm}

\usepackage{mathrsfs} 

\newcommand{\fsnull}[1]{}
\newcommand{\old}[1]{}

\usepackage[makeroom]{cancel}
\usepackage[toc,page]{appendix}
\usepackage[colorlinks=true,citecolor=blue,urlcolor=blue,linkcolor=magenta]{hyperref}

\renewcommand{\geq}{\geqslant}

\renewcommand{\vec}[1]{\boldsymbol{#1}}  

\newcommand{\bs}{\textsf{BS}}

\def\be{\begin{equation}}
\def\ee{\end{equation}}
\def\bs{\begin{split}}
\def\e{\end{split}}
\def\ba{\begin{eqnarray}}
\def\bea{\begin{eqnarray}}

\def\tea{\end{eqnarray}}
\def\ea{\end{eqnarray}}
\def\eea{\end{eqnarray}}

\newtheorem{theorem}{Theorem}

\usepackage{amssymb}
\usepackage{dsfont}

\def\be{\begin{equation}}
\def\te{\end{equation}}
\def\ee{\end{equation}}
\def\ba{\begin{eqnarray}}
\def\bea{\begin{eqnarray}}

\def\tea{\end{eqnarray}}
\def\ea{\end{eqnarray}}
\def\eea{\end{eqnarray}}

\newcommand{\beq}{\begin{equation}}
\newcommand{\eeq}{\end{equation}}

\definecolor{bluish}{rgb}{0.122, 0.435, 0.698}

\usepackage{multirow}
\usepackage{amsmath}
\usepackage{amssymb}
\usepackage{soul}
\usepackage{natbib}
\usepackage[OT4]{fontenc}
\usepackage{notes2bib}
\bibnotesetup{
note-name = ,
use-sort-key = false
}

\begin{document}
\setlength{\topmargin}{-50pt}
\setlength{\textheight}{650pt}

\title{Thresholded Quantum Sensing with a Frustrated Kitaev Trimer}

\author{C. Huerta Alderete$^\dagger$}
\affiliation{Information Sciences, CAI-3, Los Alamos National Laboratory, Los Alamos, New Mexico, USA}
\affiliation{Quantum Science Center, Oak Ridge, Tennessee 37931, USA}
\author{Anubhav Kumar Srivastava$^\dagger$}
\affiliation{Theoretical Division, Los Alamos National Laboratory, Los Alamos, New Mexico 87545, USA}
\affiliation{ICFO-Institut de Ciencies Fotoniques, The Barcelona Institute of Science and Technology,\\ Av. Carl Friedrich Gauss 3, 08860 Castelldefels (Barcelona), Spain}
\author{Bharath Hebbe Madhusudhana}
\affiliation{MPA-Q, Los Alamos National Laboratory, Los Alamos, New Mexico, USA}
\author{Andrew T. Sornborger}
\email{sornborg@lanl.gov}
\affiliation{Information Sciences, CAI-3, Los Alamos National Laboratory, Los Alamos, New Mexico, USA}
\affiliation{Quantum Science Center, Oak Ridge, Tennessee 37931, USA} \let\thefootnote\relax\footnotetext{$^\dagger$These authors contributed equally.}

\begin{abstract}
\noindent
We investigate the response of a Ramsey interferometric quantum sensor based on a frustrated, three-spin system (a Kitaev trimer) to a classical time-dependent field (signal). The system eigenspectrum is symmetric about a critical point, $|\vec{b}| = 0$, with four of the spectral components varying approximately linearly with the magnetic field and four exhibiting a nonlinear dependence. Under the adiabatic approximation and for appropriate initial states, we show that the sensor's response to a zero-mean signal is such that below a threshold, $|\vec{b}| < b_\mathrm{th}$, the sensor does not respond to the signal, whereas above the threshold, the sensor acts as a detector that the signal has occurred. 
This thresholded response is approximately omnidirectional. Moreover, when deployed in an entangled multisensor configuration, the sensor achieves sensitivity at the Heisenberg limit. Such detectors could be useful both as stand-alone units for signal detection above a noise threshold and in two- or three-dimensional arrays, analogous to a quantum bubble chamber, for applications such as particle track detection and long-baseline telescopy.
\end{abstract}
\maketitle
\vfill\eject
{\it Introduction.}
Quantum sensing, where a quantum state is prepared, allowed to interact with a signal of interest, and then measured, is used to extract information about the signal~\cite{giovannetti2006quantum, Giovannetti2011, degen2017quantum}. By using entangled input states across $N$ sensors,
one can enhance the sensitivity beyond the standard quantum limit, achieving the so-called Heisenberg limit~\cite{giovannetti2006quantum,braunstein1996geometry,braunstein1994statistical,helstrom1967minimum, Giovannetti2011, Hecht2025}. This improves the standard error from $\Delta/\sqrt{N}$ to $\Delta/N$, where $\Delta$ is the standard deviation of the signal and $N$ is the number of measurements, yielding a $\sqrt{N}$ improvement in scaling.

\begin{figure}[t!]
    \centering
    \includegraphics[scale=1]{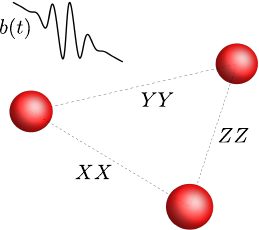}
    \caption{{\bf Kitaev trimer sensor structure.} The thresholded quantum rectifiying sensor is based on a frustrated Kitaev trimer. Here, we illustrate a magnetic field impinging on the trimer from a particular direction (black waveform). Each vertex corresponds to a spin, and couplings along the edges encode interaction strengths.}
    \label{fig:sketch}
\end{figure}

Motivated by the critical behavior of quantum materials, where properties can change rapidly with large gradients near phase transitions~\cite{Frerot2018, yang2019quantum, Liu2021, Ding2022, Hotter2024}, we introduce a minimal realization of a critical, thresholded quantum rectifying sensor, based on a frustrated three-spin Kitaev trimer (Fig.~\ref{fig:sketch}). Due to its frustrated nature, this structure exhibits behavior associated with a broken symmetry about a critical point (a crossing) in its spectrum.
Similar frustrated trimers have been studied extensively at material scales 
both theoretically and experimentally using spin-1/2 lattices~\cite{Cheng2024, Ullah2025} and spin-1 lattices~\cite{buessen2018quantum} in the presence of an external magnetic field.

Although quantum sensors are most commonly used for precise estimation of the mean phase encoded in the state of a sensor by a classical signal \cite{ramsey1950molecular,borde1989atomic,ramsey2002application,borde2002atomic,budker2007optical,balasubramanian2008nanoscale, Taylor2008,Jones2009, bal2012ultrasensitive, Boss2017, Naghiloo2017}, they can also be designed to probe higher-order moments, such as a signal’s second-order moment~\cite{schoelkopf2003qubits, MllerRigat2023}, allowing for the study of quantum fluctuations. This type of rectifying sensor can be deployed as a detector that indicates {\it whether} a signal was sensed, rather than measuring the signal amplitude itself.
One key challenge with quantum detection protocols is that, in the presence of a noisy background, the quantum detector can be rendered useless due to its sensitivity. This is particularly true for rectifying sensors, since the accrued phase increases rapidly. This Letter targets this issue. Here, we present a thresholded quantum rectifier that accumulates a phase proportional to a signal's second moment only above a threshold, $|\vec{b}| = b_\mathrm{th}$. The threshold allows the sensor to selectively measure variance in excess of the threshold that, otherwise, could rapidly push the sensor beyond the monotonic region of response (up to the first fringe) causing it to give multivalued output and rendering it unusable.

This type of quantum sensor may be used independently or in a multisensor context with an entangled input state, achieving the Heisenberg limit. In an array configuration, such sensors may be used to measure particle tracks \cite{chin2025quantum} in a similar way to the tracks measured in a bubble chamber. In photonics, this type of interferometry could be used to provide longer aperture times for quantum-enhanced long-baseline telescopy \cite{gottesman2012longer,rajagopal2024towards,aasi2013enhanced}.

{\it The thresholded rectifier.}
The thresholded rectifier (sketched in Fig.~\ref{fig:sketch}) is based on the physics of a frustrated Heisenberg–Kitaev trimer~\cite{kitaev2006anyons,trebst2022kitaev}. Without loss of generality (see the Supplemental Material~\bibnote{See Supplemental Material at \url{https://doi.org/10.1103/3z2c-2kkl} which contains additional details and proofs as well as Refs. \cite{ambainiselementary,duan2020quantum,benseny2021adiabatic,Boixo2007,Liu2015,floquetengineeringinteractionsentanglement, Zhao2023Floquet, tian2025engineeringfrustratedrydbergspin,PRXQuantum.3.020303,Omran_2019}}), the system may be described by the time-dependent Hamiltonian
\begin{equation}\label{eq:Kitaev_hamiltonian}
  H(t) =  X_1X_2 + Y_2Y_3 + Z_1Z_3 + \vec{b}(t) \cdot \vec{\sigma} \; .
\end{equation}
Here, $\vec{b}(t) = (b_X(t),b_Y(t),b_Z(t))$ is a time-dependent classical external field, and $\vec{\sigma} = (\sum_i X_i, \sum_i Y_i, \sum_i Z_i)$ is a vector of the collective operators that mediate the interaction between the classical field and the quantum sensor, where $\{X_i,Y_i,Z_i\}$ correspond to the Pauli matrices acting on the $i$th spin. The $b$-dependent spectrum arising from this Hamiltonian, for a field incident along one of the cardinal axes, is  shown in Fig.~\ref{fig:Trimer}.

In the following, we assume---supported in detail in the Supplemental Material--- that the sensor dynamics are well described by the adiabatic approximation~\cite{born1928beweis}. Starting from an initial state $|\phi_0\rangle$, the probability of the final state remaining the initial state is given by
\begin{eqnarray}
    P[\phi_0](t) &=& \vert \langle \phi_0 | U(t) | \phi_0\rangle|^2  \nonumber \\
    &=& \left|\langle \phi_0 | \Psi_0
    \exp\left[-i \int_{t_0}^t d\tau \Lambda(\tau)\right] \Psi^\dagger_0 | \phi_0 \rangle \right|^2\;,
\label{eq:AdiabaticProb}
\end{eqnarray}
where $\Lambda(t)$ is a diagonal matrix describing the eigenenergies of $H(t)$ and $\Psi_0$ is a matrix describing the eigenstates of $H(t=0)$. To use the sensor as a Ramsey interferometer, we prepare the initial state $|\phi_0\rangle$ such that $\Psi_{0}^\dagger |\phi_0\rangle = |+\rangle$, where $|+\rangle$ denotes the superposition of two eigenstates, $|+\rangle = (|\lambda_p\rangle + |\lambda_q\rangle)/\sqrt{2}$. 
Under this protocol, the probability of remaining in the initial state is given by
\begin{equation}
P[\phi_0](t) = \cos^2\left( \chi (t)\right) \; ,
\end{equation}
where we have defined the accumulated phase
\begin{equation}
    \chi (t) \equiv \frac{1}{2} \int_{t_0}^t \left(\lambda_p(\tau) - \lambda_q(\tau)\right) d\tau\; ,
\end{equation}
with $\lambda_j(t)$ denoting the instantaneous eigenenergy associated with eigenstate $\vert \lambda_j(t) \rangle$. A detailed derivation is provided in the Supplemental Material.

For $N$ entangled sensors with initial state, $\vert \Phi_+ \rangle \equiv (\ket{\lambda_p}^{\otimes N} + \ket{\lambda_q}^{\otimes N})/\sqrt{2}$, the resulting response---derived in the Supplemental Material--- is given by
\begin{equation}
  P[\Phi_+](t) = \cos^{2}\left(N\chi (t)\right) \; .
\end{equation}
This leads to a quadratic Heisenberg scaling of the quantum Fisher information, $\mathcal{F}_\chi = 4N^2$, with the variance of the accumulated phase lower bounded by the quantum Cram\'er-Rao bound of
\begin{equation}
\left(\Delta \chi (t) \right)^2 \ge \frac{1}{4{N^2}} \; .
\end{equation}
\begin{figure}[t!]

\includegraphics[scale=1]{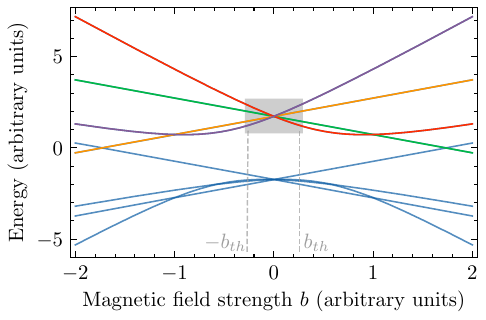}
\caption{\label{fig:Trimer} {\bf Kitaev trimer eigenspectrum.} We plot the eigenenergies $\lambda_i$ as a function of the magnetic field amplitude $b$, with the field aligned along a cardinal axis. The eigenenergies associated with a standard sensor (green and gold lines) correspond to an interferometer with support on two linear branches. In contrast, the mousetrap sensor is associated with either the (asymmetric) gold-red pair or the green-purple pair of eigenenergies. Within the gray-shaded region, where the spectrum remains approximately linear, the rectifier behaves like a standard sensor. However, when $|b|>b_\mathrm{th}$, curvature of the eigenspectrum leads to an additional phase proportional to the second moment of the signal, $b^{2}(t)$. Thus, the sensor acts as a rectifier (i.e. becomes sensitive to the signal's second moment) only above the threshold $b_\mathrm{th}$.}
\end{figure}

{\it Results.}
We now analyze the behavior of the thresholded quantum rectifier, with eigenspectrum shown in Figure~\ref{fig:Trimer}, for the example of a $b$ field in the $z$ direction. 
The eigenenergies extend outward from two gapped critical points at energies of $\pm \sqrt{3}$ at $|\vec{b}| = 0$. Our focus will be on the upper critical point, where four spectral lines emerge. Two of these lines (shown in gold and green) are linear in the $b$ field. If the initial state $|+\rangle$ has support only on the eigenstates corresponding to these spectral lines, the sensor behaves as a standard sensor that measures the mean of the $b$ field, i.e.,
\begin{equation}
  P[\phi_0](t) = \cos^{2}\left(N \chi_\mathrm{std}(t)\right) = \cos^{2}\left(N\int_{t_0}^t c\; b(\tau) d\tau\right) \; ,
\end{equation}
where $c$ characterizes the slope of the linear dependence.

In contrast, if the initial state has support on either the green-purple or gold-red pairs of spectral lines depicted, the sensor response differs. For small field amplitudes, $|\vec{b}|<b_\mathrm{th}$, indicated by the gray-shaded region, the eigenspectrum remains approximately linear and the sensor effectively measures the mean. However, above this threshold, $|\vec{b}|>b_\mathrm{th}$, the linear symmetry of the eigenspectrum breaks to the left for the green-purple pair and to the right for the gold-red pair, and the eigenenergies become nonlinear in $b$. In this regime, the accumulated phase becomes approximately proportional to $b^2(t)$, as shown in the Supplemental Material.  The device therefore acts as a thresholded rectifier, detecting that the field has exceeded $b_\mathrm{th}$. 
These two pairs of spectral lines result in what we call {\em quantum mousetrap sensors}.

\begin{figure}[t!]
\includegraphics[scale=1]{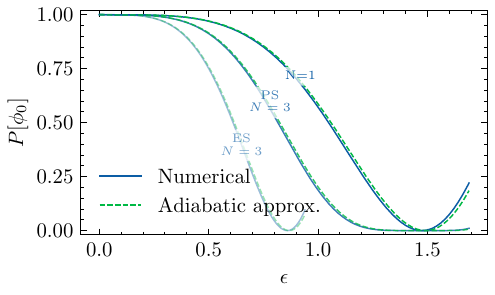}
\caption{\label{fig2} {\bf Trimer mousetrap sensor response.} The response probability, $P[\phi_0]$, shown as a function of the signal amplitude, $\epsilon$, for a zero-mean, oscillatory signal with Gaussian envelope, $b(t) = \epsilon \exp(-t^{2}/s) \sin(2\pi k t)$, where $k = 0.05$, $s = 20$, $t_0 = -15$, and $t = 15$. Results are shown for $N=1$ and $N=3$ to illustrate how the response scales with the number of sensor copies. Both product state (PS) and entangled state (ES) are used as inputs for $N=3$. Note that a response is not incurred until $\epsilon$ is large enough ($\approx 0.3$) to be affected by the change in curvature of $\lambda_i$ near the asymmetric wells in the spectrum, with basins centered at $b = \pm 0.8$ (see Fig.~\ref{fig:Trimer}).}
\end{figure}
\begin{figure}[t!]
\includegraphics[scale=1]{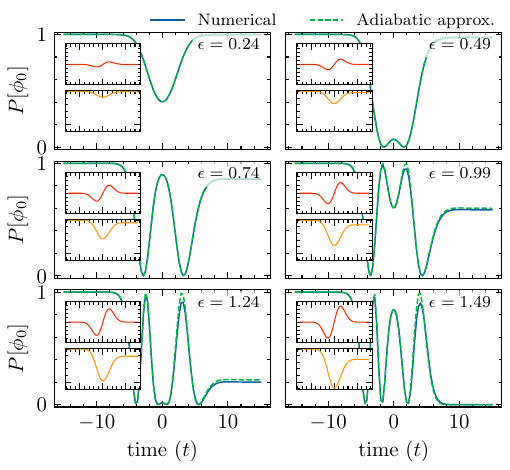}
\caption{\label{fig3} {\bf Mousetrap sensor response.} The probability, $P[\phi_0]$, is shown as a function of time, $t$, for a range of signal amplitudes, $\epsilon$, for a zero-mean, oscillatory signal with Gaussian envelope, $b(t) = \epsilon \exp(-t^{2}/s) \sin(2\pi k t)$, where $k = 0.05$, $s = 20$, $t_0 = -15$, and $t = 15$. Insets show the signal $b(t)$ (top left in red) and the accumulated phase $\chi_\mathrm{mt}(t)$ (lower left in orange), both plotted on the same time scale as the main figure, with vertical ranges of $(-1,1)$ and $(-6,0)$, respectively. For small amplitudes ($\epsilon \lesssim 0.3$), the signal remains within the approximately linear region of the potential well, resulting in minimal phase accumulation and no integrated response. As $\epsilon$ exceeds the threshold near $0.3$, portions of the signal probe the nonlinear regime, where symmetry breaking in the spectrum leads to phase accumulation. The sensor thus acts as a signal rectifier responding only when the signal amplitude crosses $b_\mathrm{th}$, where it begins to integrate deviations from linearity.}
\end{figure}

The response probability for the mousetrap sensor is $P[ \phi_{0} ] = \cos^{2} \left( N\chi_\mathrm{mt}(t) \right)$, where 
\begin{eqnarray}
  \chi_\mathrm{mt}(t) & = & \frac{1}{2}\int_{t_{0}}^{t} \left(\sqrt{3} + 2b(\tau) -\sqrt{3 - 4b(\tau) + 4b^{2}(\tau)}\right) d\tau \; .\nonumber\\
  \label{eq:chimt}
\end{eqnarray}
In Fig.~\ref{fig2}, we plot $P[ \phi_{0} ]$ for a zero-mean, oscillatory $b$ field with Gaussian envelope.
Three response probabilities are shown, one for $N=1$, a single spin, one for $N=3$ with product state input giving the standard quantum limit, and one for $N=3$ with entangled state input giving the evident Heisenberg limit.

The mousetrap sensor response over the range $\epsilon \in (0, 1.49)$ is shown in Fig.~\ref{fig3}. The plateau observed  below the threshold at $\epsilon \sim 0.3$ in Fig.~\ref{fig2} can be understood by examining the integral, $\chi_\mathrm{mt}$ (Fig. 4, inset of each panel), in Eq.~\ref{eq:chimt}. When the signal probes only the approximately linear region of the spectrum (highlighted by the gray box in Fig.~\ref{fig:Trimer} about the point $(0,0)$, and depicted in Fig. ~\ref{fig3}, top-left panel), the zero-mean signal oscillates back and forth from positive to negative values, with $P[\phi_0]$ averaging to zero, giving no accumulated response. In contrast, for $\epsilon \gtrsim 0.3$, the eigenenergies deviate from linearity, and the signal accrues. Thus, the mousetrap sensor functions as a thresholded rectifier, sensitive to signal amplitudes in the approximate range $\epsilon \in [0.3,1.5]$ with the maximum (first fringe) response near $\epsilon \sim 1.5$. This operating window is stated in dimensionless units; see the Supplemental Material for the nondimensionalization that maps physical field amplitudes and pulse durations.

Although the adiabatic approximation can fail near degeneracies, the present regime is well described within the adiabatic treatment used here. Near the critical, point the spectrum becomes approximately linear and admits a closed-form adiabatic description. Consequently, the adiabatic approximation remains highly accurate for the mousetrap protocol, as verified numerically by comparison to full time-dependent simulations (see Figs.~\ref{fig2} and \ref{fig3} and the Supplemental Material). We have additionally shown that more realistic Ornstein-Uhlenbeck processes with appropriate bandwidth also exhibit mousetrap behavior (see the Supplemental Material).

\begin{figure}
    \centering
    \includegraphics[scale=1]{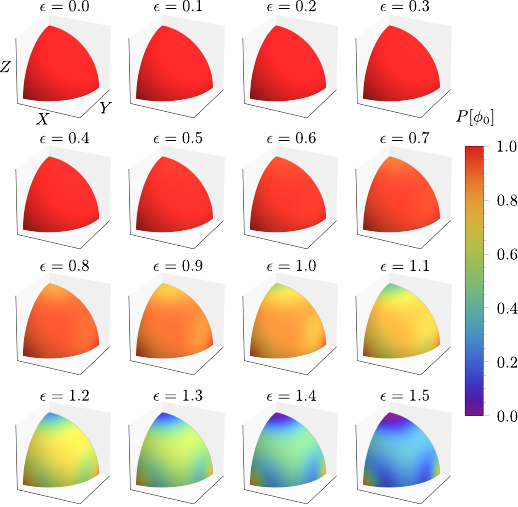}
    \caption{{\bf Omnidirectional mousetrap threshold response}. This figure shows the probability $P[\phi_0]$ of a mousetrap remaining in its initial state as a function of the incident direction of the $b$ field, using the same Gaussian signal and parameters as in Figs.~\ref{fig2} and \ref{fig3}. Due to the eight-fold symmetry of the response, only one octant of the sphere is shown. The response color mapping is indicated by the legend. At low amplitudes ($\epsilon \lesssim 0.3$), the sensor remains unresponsive across all directions (red), consistent with the plateau observed in Fig.~\ref{fig2}. As $\epsilon$ increases, directional sensitivity begins to emerge around $\epsilon \approx 0.7$ (orange, yellow, green, blue), with significant anisotropic variations appearing in the range $0.7 < \epsilon < 1.5$, where the signal begins to probe the nonlinear regions of the spectrum.}
    \label{fig:mousetrap-incidence}
\end{figure}

To assess directional response, we numerically studied the mousetrap's behavior as a function of magnetic field incidence angle for fixed initial state (with a sensor initialized to eigenstates optimal for incidence along the $z$ axis, but tested for all incidence directions in the octant). As shown in Fig.~\ref{fig:mousetrap-incidence}, while specific parameter values vary for different incidence angles, the mousetrap consistently exhibited the same response to $b$ fields at all angles: a plateau in probability for small signal amplitudes ($\epsilon < 0.3$) followed by a significant response from $0.3 < \epsilon < 1.5$, consistent with Fig.~\ref{fig2}. At higher amplitudes ($\epsilon \approx 0.7$) the response becomes increasingly sensitive to the incident angle, still, there is always a linear region around the critical point giving rise to the mousetrap thresholding effect. As a result, the quantum mousetrap can be used as an omnidirectional, thresholded rectifier and, hence, can be used in a two- or three-dimensional array to detect incident $b$ fields in all directions. This frustration-induced omnidirectionality is a practical advantage of the trimer architecture, since the thresholded sensing characteristics persist under variations in the signal direction.

Many different combinations of the trimer eigenspectra may be used to construct quantum sensors. Of these, two are thresholded---the mousetrap and an additional sensor with
\begin{eqnarray} \nonumber
  \chi_\mathrm{mt}(t) & = & \frac{1}{2}\int_{t_{0}}^{t} \left(-2 b(\tau) - \sqrt{3 - 4 b(\tau) + 4 b^2(\tau)} \right.\\
  && \qquad \qquad
  \left. + \sqrt{3 + 4 b(\tau) + 4 b^2(\tau)} \right) ~ d\tau \; ,
\end{eqnarray}
which may be used as a thresholded sensor of the third moment of $\vec{b}(t)$ (the skew, from purple and red eigenstates in Fig.~\ref{fig:Trimer}).

Due to their complexity, quantum mousetraps can be difficult to implement. In the Supplemental Material, we outline an experimental protocol for realizing a mousetrap using a neutral-atom (Rydberg) platform. In brief, the protocol involves initializing a linear array of $3N$ qubits prepared in the standard Greenberger–Horne–Zeilinger (GHZ) state,  $\vert GHZ \rangle = (\ket{0}^{\otimes 3N} + \ket{1}^{\otimes 3N})/\sqrt{2}$; rearranging the $3N$ qubits into trimers. And, finally, quenching the system to $H(p=0)=H_{\text{init}}$ and adiabatically sweeping the parameter $p$ to $1$. This prepares the desired GHZ-like state. Finally, as in most GHZ-based sensing protocols, the preparation sequence is reversed after the phase-acquisition stage so that the accumulated phase can be read out in the computational basis.
This implementation perspective naturally suggests a modular route to scaling, in which the trimer acts as a minimal sensing unit that can be replicated to build an $N$-trimer sensor. Entanglement resources (e.g., GHZ-type states across trimers) then provide a direct route to quantum-enhanced scaling, with the main additional challenge being preparation of the required GHZ-like eigenstate superposition across trimers.

{\it Discussion.}
The thresholded quantum rectifier and the additional sensors outlined in the Supplemental Material may be combined to measure quantities of interest for the analysis of the input $b$ field. For instance, we can use a standard sensor and a Landau-Zener sensor to measure the first ($\int_{t_0}^t b(\tau) d\tau = \arccos{\sqrt{P_\mathrm{std}[+]}}$) and second ($\int_{t_0}^t b^2(\tau) d\tau = \arccos{\sqrt{P_\mathrm{LZ}[+]}}$) moments, respectively, to postcompute the variance $\mathrm{Var}[b(t)] = \int_{t_0}^t b^2(\tau) d\tau - \left|\int_{t_0}^t b(\tau) d\tau\right|^2$. Similarly, for a thresholded rectifier, if the mean is nonzero, it can be separately measured with a standard sensor and subtracted.

There has been recent interest in making use of quantum computers to postprocess quantum data acquired by quantum sensors~\cite{zhou2018achieving,chin2025quantum,allen2025quantum}. Since they are coherent, mousetraps and other spin systems, such as those in the Supplemental Material, could be considered as quantum sensing modules that preprocess the quantum data before incorporation in larger quantum computational systems, such as quantum phase estimation~\cite{allen2025quantum,patel2024optimal}. From this point of view they are a first example of quantum computation at the edge \cite{allen2025quantum, khan2025quantum,khan2025quantumcomputational}.

{\it Acknowledgments}--- We thank Arnab Banerjee, Daniel Bowring, David Weld, Tyler Volkoff, and Guillem Müller-Rigat for helpful discussions. A.K.S. was a participant in the 2024 Quantum Computing Summer School at LANL. This work was supported by the U.S. Department of Energy (DOE) through a quantum computing program sponsored by the Los Alamos National Laboratory (LANL) Information Science \& Technology Institute. This material is based upon work supported by the U.S. Department of Energy, Office of Science, National Quantum Information Science Research Centers, Quantum Science Center (QSC). A.T.S. was funded by the QSC during the conception of the project. A.T.S. and C.H.A. acknowledge support for a numerical feasibility study of nonlinear quantum sensors from LANL's LDRD Program under project 20248110CT-IMS. Subsequent numerical and analytical work was performed via the QSC. C.H.A. was also supported by the U.S. Department of Energy, Office of Science, Office of Advanced Scientific Computing Research, under the Computational Partnerships program. A.K.S. acknowledges support from the European Union’s Horizon 2020 Research and Innovation Programme under the Marie Sklodowska-Curie Grant Agreement No. 847517; MCIN/AEI (PGC2018–0910.13039/501100011033, CEX2019-000910-S/10.13039/501100011033, Plan National STAMEENA PID2022-139099NB, project funded by MCIN and by the “European Union NextGenerationEU/PRTR” (PRTR-C17.I1), FPI); Ministry for Digital Transformation and of Civil Service of the Spanish Government through the QUANTUM ENIA project call ---Quantum Spain project, and by the European Union through the Recovery, Transformation and Resilience Plan--- NextGenerationEU within the framework of the Digital Spain 2026 Agenda; CEX2024-001490-S [MICIU/AEI/10.13039/501100011033]; Fundació Cellex; Fundació Mir-Puig; Generalitat de Catalunya (European Social Fund FEDER and CERCA program); Barcelona Supercomputing Center MareNostrum (FI-2023-3-0024); European Union HORIZON-CL4-2022-QUANTUM-02-SGA – PASQuanS2.1, 101113690; EU Horizon 2020 FET-OPEN OPTOlogic, Grant No 899794; QU-ATTO, 101168628; EU Horizon Europe research and innovation program under grant agreement No. 101080086 NeQST. This paper is approved by Los Alamos National Laboratory for universal release under designation LA-UR-25-24691. 

\bibliography{local.bib}

\onecolumngrid
\newpage
\appendix
\noindent

\section*{Supplemental Material for {\em ``Thresholded Quantum Sensing with a Frustrated Kitaev Trimer"}}

In this Supplemental Material, we provide additional details for the manuscript {\em Thresholded Quantum Sensing with a Frustrated Kitaev Trimer}. We begin by outlining the non-dimensionalization procedure used to cast the sensor Hamiltonians into a standard form (Sec. \ref{app:Nondim}), followed by a discussion of the adiabatic approximation and its regime of validity for the Kitaev trimer sensor (Sec. \ref{app:Adiabatic}). In Sec. \ref{app:complexspectra}, we derive the accumulated phase for arbitrary pairs of eigenstates, which governs the Ramsey interferometric response. Section \ref{app:Heisenberg} generalizes the sensor protocol to multi-device configurations, establishing the conditions under which Heisenberg-limited scaling is achieved. Sec. \ref{app:TrimerSensor} details the spectral properties of the trimer and provides explicit expressions for the accumulated phase in the thresholded sensing regime. Here, we additionally investigate the response to both smooth deterministic and, more relevant for experiment, to stochastic Ornstein-Uhlenbeck processes. Finally, Sec.~\ref{app:Experimental} gives detailed rationale and specifics for an experimental implementation of a quantum mousetrap.

\section{Non-Dimensionalization of Sensor Hamiltonians}\label{app:Nondim}
A sensor Hamiltonian, $H(t)$, may be described as
\begin{equation}
  H(t) = \alpha H_0 + \beta(t) H_1 \; ,
\end{equation}
where $\alpha$ is the strength of the bare Hamiltonian, $H_0$, and $\beta(t)$ is a time-dependent function that modulates the interaction Hamiltonian, $H_1$. Thus, the Schr\"odinger equation becomes
\begin{equation}
  \frac{\partial}{\partial t} |\psi\rangle = -i \left(\alpha H_0 + \beta(t) H_1\right) |\psi\rangle \; .
\end{equation}
Dividing by $\alpha$, and introducing a new time variable, $\alpha t \equiv t'$, gives
\begin{equation}
  \frac{\partial}{\partial t'} |\psi\rangle = -i \left(H_0 + \frac{\beta(t'/\alpha)}{\alpha} H_1\right) |\psi\rangle \; .
\end{equation}
Defining $\beta(t'/\alpha)/\alpha \equiv b(t')$, then dropping the prime, we now have
\begin{equation}
  \frac{\partial}{\partial t} |\psi\rangle = -i \left(H_0 + b(t) H_1\right) |\psi\rangle \; .
\end{equation}
This form is applicable to all sensor Hamiltonians discussed in this manuscript by appropriate rescaling of both time, $t$, and the physical field, $\beta$.

\section{The Adiabatic Approximation}\label{app:Adiabatic}

We begin this Section with a Theorem used below. In the following, we assume that $H(t)$ is such that the adiabatic theorem applies. That is, from~\cite{ambainiselementary}, we have (verbatim) 
\begin{theorem}
Let $H(s)$, $0 \le s \le 1$, be a time-dependent Hamiltonian. Let $\vert \lambda(s) \rangle$ be one of its eigenstates with corresponding eigenvalue $\lambda(s)$. Assume that for any $s \in [0, 1]$, all other eigenvalues of $H(s)$
are either smaller than $\lambda(s) - \gamma$ or larger than $\lambda(s)+\gamma$ (i.e., there is a spectral gap of $\gamma$ around $\lambda(s)$). Consider the adiabatic evolution given by $H$ and $\vert \lambda \rangle$ applied for time $T$. Then, the following condition is
enough to guarantee that the final state is at most a distance $\delta$ from $\vert \lambda(1)\rangle$:
\begin{equation}\label{eq:theorem}
  T \ge \frac{10^5}{\delta^2} · \max\left\{
\frac{\|H'\|^3}{\gamma^4}, \frac{\|H'\|\cdot \|H''\|}{\gamma^3}\right\} \; .
\end{equation}
\label{thm:quantumadiabatic}
\end{theorem}
Therefore, as long as $H$ has a spectral gap $1/\mathrm{poly}$ around $\lambda$, we can reach a state that
is at most $1/\mathrm{poly}$ away from $\vert \lambda(1)\rangle$ in polynomial time. We note that the $\gamma$ dependence can be improved for this theorem~\cite{duan2020quantum}. 

Additionally, we note that for the quantum mousetrap sensor that is this Letter's focus, the eigenspectrum becomes approximately linear near the critical point. As a result, in this region of crossing eigenspectra, the adiabatic approximation holds exactly due to linearity. Consequently, the adiabatic theorem applies broadly to the full evolution of the sensor response~\cite{benseny2021adiabatic}. We have tested this numerically (see Figs.~3 and 4 of the main text).

For a system described by a Hamiltonian, $H(t)$, the time evolution unitary can be expressed as a time-ordered exponential:
\begin{equation}\label{eq:TimeOrderedEvolution}
  U(t) = \mathcal{T} \exp \left[-i \int_{t_{0}}^{t} d\tau H(\tau)\right] \; 
\end{equation}
where $\mathcal{T}$ denotes the time-ordering. 
This may be rewritten as the limit of a Riemann sum
\begin{equation}
U = \mathcal{T} \exp \left[-i \lim_{\Delta\tau \rightarrow 0}\sum_{n=0}^{k} \Delta\tau H(\tau_n)\right] \; .
\end{equation}
This expression can then be written as an ordered product of exponentials. We have
\begin{equation}
  U = \lim_{\Delta\tau \rightarrow 0}\prod_{n=k}^0 \exp(-i \Delta\tau H(\tau_n)) \; .
\end{equation}

Using the diagonalization of the Hamiltonian, $H(t_n) = \Psi_n \Lambda_n \Psi^\dagger_n $, where $\Psi_{n}$ and $\Lambda_{n}$ are the instantaneous eigenstate and eigenenergy matrices at time $t_n$, respectively, and assuming that the adiabatic approximation holds, ({\it i.e.} the eigenvector matrices, $\Psi_n$, vary slowly with $n$ such that $\Psi_{n+1}^\dagger \Psi_n \sim I$), we have:
\begin{eqnarray}
 U &=& \lim_{\Delta\tau \rightarrow 0} \prod_{n=k}^0 \Psi_n \exp(-i \Delta\tau \Lambda_n) \Psi^\dagger_n \; , \nonumber \\
  &\approx& \lim_{\Delta\tau \rightarrow 0} \Psi_k \left[ \prod_{n=k}^0 \exp(-i \Delta\tau \Lambda_n) \right] \Psi^\dagger_0 \; .
\end{eqnarray}

This expression allows us to rewrite the evolution operator as a diagonalization with eigenstate matrices, $\Psi_t$, sandwiching a diagonal eigenenergy matrix, $\Lambda(t)$, that is an exponential of a sum and, taking the continuous limit, we recover the integral form
\begin{eqnarray}
    U &\approx& \lim_{\Delta\tau \rightarrow 0} \Psi_k \exp\left[-i \sum_{n=0}^k \Delta\tau \Lambda_n\right] \Psi^\dagger_0 \; \nonumber \\
  &=& \Psi_t \exp\left[-i \int_{t_0}^t d\tau \Lambda(\tau)\right] \Psi^\dagger_0 \;.
  \label{eq:AdiabaticApprox}
\end{eqnarray}

We now assume that the $b$-field is initially zero and non-zero only during an exposure time $T^\prime < T=t-t_{0}$ after an initial time, $t_0$. This implies $\Psi_{t_{f}} = \Psi_{t_{0}}$. 
With this expression in hand, we determine the final state probability for a system to remain in its initial state, $|\phi_0\rangle$,
\begin{eqnarray}
    P[\phi_{0}](t) &\approx& | \langle \phi_0 | U(t) | \phi_0\rangle|^2 \nonumber \\
   &=& \left|\langle \phi_0 | \Psi_0 \exp\left[-i \int_{t_0}^t d\tau \Lambda(\tau)\right] \Psi^\dagger_0 | \phi_0 \rangle\right|^2\;. \nonumber \\
  \label{eq:AdiabaticProb}
\end{eqnarray}

As mentioned in the main text, to use the sensor as a Ramsey interferometer, we choose the initial state $|\phi_0\rangle$ such that $\Psi_{0}^\dagger |\phi_0\rangle = |+\rangle$, where $|+\rangle$ denotes the normalized superposition of two eigenstates, $|+\rangle = (|\lambda_p\rangle + |\lambda_q\rangle)/\sqrt{2}$. Combined with the fact that $\Psi_{t_{f}} = \Psi_{t_{0}}$, this leads to $\langle \phi_0| \Psi_{t_{f}} = \langle +|$.

\begin{figure}
    \centering
    \includegraphics[scale=1]{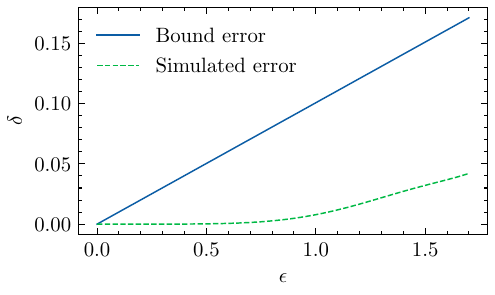}
    \caption{{\bf Adiabatic approximation error.} The adiabatic approximation error computed numerically (dashed green) compared to the lower bound  \eqref{eq:accuracy} (solid blue) on the accuracy $\delta$ shown as a function of the signal amplitude $\epsilon$ for a zero-mean, oscillatory signal with Gaussian envelope, $b(t) = \epsilon \exp(-t^{2}/s) \sin(2\pi k t)$, where $k = 0.05$, $s = 20$, $t_0 = -15$, and $t = 15$.}
    \label{fig:accuracy}
\end{figure}

To guarantee adiabaticity with final state error bounded by $\delta$, the runtime must satisfy Theorem \ref{thm:quantumadiabatic}. In order to understand how the adiabatic approximation applies to a mousetrap, we reparametrize time using $t= T s - t_{f}$, with $s\in [0,1]$, to standardize the time interval. Under this transformation, the first and second time derivatives of the Hamiltonian scale as $H^{\prime} =  T b^{\prime} (t) H_{1}$ and $H^{\prime \prime} =  T^{2} b^{\prime \prime} (t) H_{1}$, leading to operator norms $\|H'\|= \vert 3 T b^{\prime}(t) H_{1} \vert$ and $ \|H''\| = \vert 3 T^{2} b^{\prime}(t) H_{1} \vert$. Substituting these expressions into the adiabatic condition yields a lower bound for $\delta^{2}$ in terms of the runtime, $T$, the spectral gap, $\gamma$, and the derivatives of the magnetic field, $b(t)$, 
\begin{equation}\label{eq:accuracy}
    \delta^{2} \geq \frac{10^5}{T} · \max\left\{
\frac{\vert 3 T b'(t) |^3}{\gamma^4}, \frac{| 9 T^{3} b'(t) b''(t)|}{\gamma^3}\right\} \; .
\end{equation}
For the set of parameters used in Fig.~2 of the main text and with $\epsilon\sim1.5$, we estimate a lower bound for $\delta^2$, $\delta^{2} \gtrsim 2.28 \times 10^{-2}$, or $\delta \gtrsim 0.15$. In Fig. \ref{fig:accuracy}, we compare the numerical error of the adiabatic approximation, Eq. \eqref{eq:AdiabaticApprox}, that is 
\begin{equation}
\delta = 1 - \vert \langle\psi_{numeric} \vert \psi_{adiabatic} \rangle \vert^{2} 
\end{equation}
with $\delta$ computed from Eq. \eqref{eq:accuracy}. The results show that the bound of Thm.~\ref{thm:quantumadiabatic} is somewhat loose. This behavior of the approximation is not unexpected and has been pointed out in the literature \cite{benseny2021adiabatic}. Additionally, as we mentioned above and in the main text, for $b$-fields at small amplitudes about the critical point, the approximation becomes essentially exact as the system approaches linearity at the crossing. This behavior is readily seen in Fig.~\ref{fig:accuracy} for $\epsilon \lesssim 0.6$, where $\delta \approx 0$.

\section{Computing $\chi(t)$ for Arbitrary Pairs of Eigenstates}\label{app:complexspectra}
By carefully selecting the initial state $\vert \phi_0 \rangle$ to have support on only two of the instantaneous eigenstates, $\vert \lambda_p \rangle$ and $\vert \lambda_q \rangle$, the sensor will function as a Ramsey interferometer and exhibit a cosinusoidal modulation in its response probability. To make this relationship explicit, we define the sum, $S$, and difference, $D$, of the two eigenenergies (suppressing their explicit time dependence for now),
\begin{eqnarray}
  S & = & \frac{1}{2}(\lambda_p + \lambda_q) \nonumber\\
  D & = & \frac{1}{2}(\lambda_p - \lambda_q) \; .
\end{eqnarray}
In this two-level subspace, the system accumulates a global phase proportional to $S$, which does not affect the measurement outcome, and a relative phase, proportional to $D$, that gives rise to the $b$-field dependence in the response probability,
\begin{eqnarray}\label{eq:response}
  P_{p/q}[+](t) & = & \left|\frac{1}{2} \left(e^{-i \int_{t_0}^t\lambda_p(\tau)d\tau}  +  e^{-i \int_{t_0}^t\lambda_q(\tau)d\tau}\right) \right|^2 \nonumber\\
  & = & \left| e^{-i\int_{t_0}^t S(\tau) d\tau} \cos\left(\int_{t_0}^t D(\tau) d\tau\right)\right|^2  \nonumber\\
  & = &  \cos^{2}\left(\int_{t_0}^t D(\tau) d\tau\right) \nonumber\\
  & \equiv & \cos^{2}(\chi_{p/q}(t))\; .
\end{eqnarray}
Considering all possible combinations of instantaneous eigenvalues and approximating with a power series for small values of $b(t)$, a general expression for this type of sensor response probability is given by 
\begin{equation}
    P_{p/q}[\phi_{0} ](t) = \cos^{2} \left( \sum_{i} \alpha_{i} f_{i} (t)\right),
\end{equation}
where the $\alpha_{i}$ are constants that depend on the chosen pair of eigenvalues and $f_{i}(t) = \int_{t_{0}}^{t} b^{i}(\tau)d\tau$ indicates the moments of the $b$-field. 
\section{Multi-Device Sensing and the Heisenberg Limit}\label{app:Heisenberg}
Consider a multi-sensor with Hamiltonian
\begin{equation}
  H = H^s_1 + H^s_2 + ... + H^s_N \; ,
\end{equation}
where $H^s_i$ denotes the $i$'th sensor Hamiltonian, $H^s$, and $[H_{i}, H_{j}] = 0, ~ \forall~ i,j$. For each sensor, we have the unitary evolution given in Eq.~\ref{eq:TimeOrderedEvolution}. This results in a unitary describing the multi-sensor evolution $\mathcal{U}(t) = U(t)^{\otimes N} \;.$
To obtain the standard quantum limit, we take $N$ copies of the initial state
\begin{equation}\label{eq:prod_state}
  \vert \Phi_0 \rangle \equiv |\phi_0\rangle^{\otimes N}
\end{equation}
giving the time-dependent state probability
\begin{align}
P[\Phi_0](t) &= |\langle \Phi_0 |  \left(\Psi_0 \exp\left[-i \int_{t_0}^t d\tau \Lambda(\tau)\right]\Psi^\dagger_0 \right)^{\otimes N}|\Phi_0\rangle|^2 \; . \nonumber \\ 
\end{align}

Considering an initial state that is a superposition of two eigenstates, the probability of finding a multi-sensor system in the $\Psi_0^{\dagger\ \otimes N} \vert \Phi_0 \rangle \equiv |+\rangle^{\otimes N}$ state is given by
\begin{eqnarray}\label{eq:probNtrimers}
  P_{p/q}[\Phi_0](t) & = & \left\vert \cos^2\left( \chi(t) \right) \right|^N \nonumber\\ & = & \cos^{2N}\left(\frac{1}{2} \int_{t_0}^t \left(\lambda_p(\tau) - \lambda_q(\tau)\right) d\tau \right)\; .
\end{eqnarray}

The Fisher information (FI) is given by, 
\begin{equation}\label{eq:FI_prob}
    \mathcal{F}_{\chi}=\frac{(\partial_\chi P_{p/q}[\Phi])^2}{P_{p/q}[\Phi](1-P_{p/q}[\Phi])}\; ,
\end{equation} for a projective measurement on the state $\vert \Phi \rangle$. For a system with $N$-trimers following Eq.~\eqref{eq:probNtrimers}, it can be verified that the FI scales with $N$ owing to its additive nature, recovering the shot-noise scaling~\cite{braunstein1994statistical, giovannetti2006quantum, Boixo2007, Liu2015}. 

To obtain the Heisenberg limit, we consider an entangled input of GHZ-type
\begin{equation}\label{eq:ghz_type}
    |\Phi_+\rangle \equiv \frac{1}{\sqrt{2}} \left(|\lambda_{p}\rangle^{\otimes N} + |\lambda_q\rangle^{\otimes N} \right)\; .
\end{equation}
The phase difference in the integral of Eq.~\eqref{eq:response} scales with $N$, giving the time-dependent state probability as
\begin{equation}\label{eq:Heis}
    P_{p/q}[\Phi_+](t) = \left\vert \cos^2\left( N\chi(t) \right) \right|\;.
\end{equation}
From Eq.~\eqref{eq:FI_prob}, for entangled input, we recover the Heisenberg scaling of the FI given by $\mathcal{F}_\chi= 4N^2$. We emphasize that we do not carry out a full parameter-estimation analysis in this work; instead we characterize the sensing capability through the response probability and its scaling with quantum resources.

\begin{figure*}[t]
     \centering
     \includegraphics[scale=1]{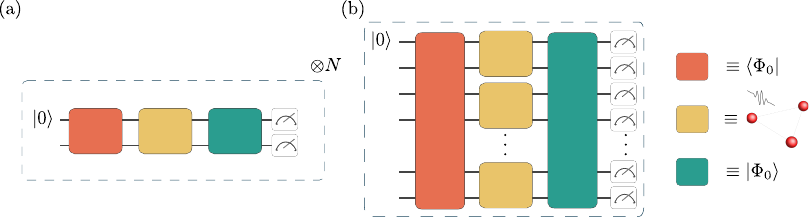}
     \caption{{\bf Multi-sensor scaling.} Multiple copies of the trimer sensor are prepared either in (a) the product state $\vert \Phi_0 \rangle \equiv |\phi_0\rangle^{\otimes N}$  or (b) in an entangled GHZ-type state $|\Phi_0\rangle = |\Phi_+\rangle \equiv \frac{1}{\sqrt{2}} \left(|\lambda_{p}\rangle^{\otimes N} + |\lambda_q\rangle^{\otimes N} \right)$. Interactions with the signal (red nodes) generate phase accumulation across the sensors. Product states yield standard quantum-limit scaling, while entangled states can achieve Heisenberg scaling by coherently enhancing the accumulated phase. }
     \label{fig:scaling}
 \end{figure*}

In Figure \ref{fig:scaling}, we depict these two possible types of multi-sensor use. Multiple copies of the trimer sensor can be prepared either in product states, Eq. \eqref{eq:prod_state}, or in an entangled GHZ-type configuration, Eq. \eqref{eq:ghz_type}. In the product-state case, each copy accumulates phase independently, and the joint survival probability is simply the product of single-sensor contributions, yielding the standard quantum limit. In contrast, in the entangled configuration the accumulated phase is coherently amplified by a factor of $N$, achieving the quadratic Heisenberg scaling described in Eq. \eqref{eq:Heis}.

\section{Measuring a Thresholded 2nd Moment of a Signal with a Kitaev Trimer Sensor Exhibiting a Broken Symmetry}\label{app:TrimerSensor}

Finally, we turn to a complete description of the analysis of the {\em quantum mousetrap} sensor, a three-qubit system based on the frustrated trimer Hamiltonian. As discussed in the main text, this three-qubit sensor has interesting signal analysis properties. It may be used as a thresholded rectifier or to measure a thresholded third moment (skew). 

Due to its cartesian symmetry, one can choose a magnetic field along any axis. Here, we choose the $z$-axis, giving the Hamiltonian, $H(t)= X_{1}X_{2} + Y_{2}Y_{3} + Z_{1}Z_{3} + b(t) \sum_{i=1}^{3}Z_{i}$. This Hamiltonian is diagonalizable with $b$-field-dependent eigenvalues and eigenvectors. An approximate solution is obtained via the adiabatic approximation (Eq.~\ref{eq:AdiabaticProb}).
The instantaneous eigenvalues are given by 
\begin{eqnarray}
  \Lambda(t) & = & \mathrm{diag}( \lambda_1, \lambda_2, \lambda_3, \lambda_4, \lambda_5, \lambda_6, \lambda_7, \lambda_8 ) \nonumber\\ \nonumber
  & = & \mathrm{diag}( -p - s, - p + s, p - s, p + s, \\ 
  && \qquad -p - q, -p + q, p - r, p + r )
\end{eqnarray}
where $s = \sqrt{3}$, $p = b(t)$, $q=\sqrt{3 - 4 b(t) + 4 b(t)^2}$ and $r = \sqrt{3 + 4 b(t) + 4 b(t)^2}$. The corresponding matrix of (normalized) eigenvectors evaluated at $t=0$ is given by 
\begin{eqnarray}
\Psi_{0} = \left(
\begin{array}{cccccccc}
 0 & 0 & 0 & a & 0 & a-b &
   a & 0 \\
 0 & 0 & 0 & c & 0 & c+d &
   c & 0 \\
 0 & a & a-b & 0 &
   a & 0 & 0 & 0 \\
 0 & c & c+d & 0 &
   c & 0 & 0 & 0 \\
 0 & -a & 0 & 0 & a & 0 & 0 & e \\
 0 & c & 0 & 0 & -c & 0 & 0 & h \\
 f-g & 0 & 0 & -f & 0 & 0 &
   f & 0 \\
 c+d & 0 & 0 & -c & 0 & 0 &
   c & 0 \\
\end{array}
\right), \nonumber \\
\end{eqnarray} with constants
\begin{align}
a &= \frac{1}{\sqrt{2+\left(\sqrt{3}-1\right)^2}}, &
b = \sqrt{\frac{3}{2+\left(\sqrt{3}-1\right)^2}}, \\ \nonumber
c &= \frac{1}{\sqrt{2+\left(1+\sqrt{3}\right)^2}}, &
d = \sqrt{\frac{3}{2+\left(1+\sqrt{3}\right)^2}} \\ \nonumber
e &= \frac{\sqrt{3}-1}{\sqrt{2+\left(\sqrt{3}-1\right)^2}}, &
f = \frac{1}{\sqrt{2+\left(1-\sqrt{3}\right)^2}}, \\ \nonumber 
g &= \sqrt{\frac{3}{2+\left(1-\sqrt{3}\right)^2}}, & 
h = \frac{1+\sqrt{3}}{\sqrt{2+\left(1+\sqrt{3}\right)^2}}
\end{align}
 \begin{figure*}[t]
     \centering
     \includegraphics[scale=1]{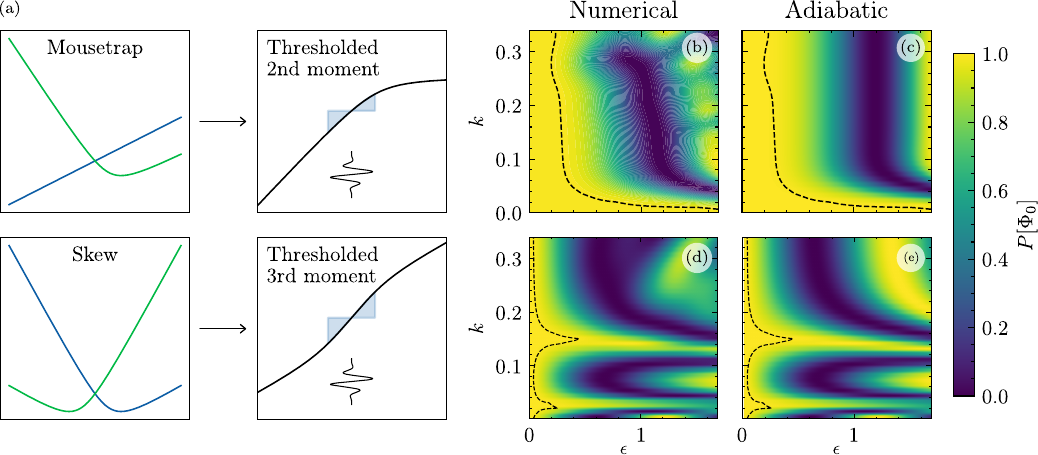}
     \caption{{\bf Thresholded sensors.} Construction of the mousetrap (top) and skew (bottom) sensors from pairs of trimer eigenenergies. (a) Left panels: selected energy pairs, $\lambda_4$ and $\lambda_6$ (top, mousetrap) and $\lambda_6$ and $\lambda_8$ (bottom, skew). Right panels: corresponding instantaneous phase difference (shown in black) and accumulated phases for the pulse (inset at bottom of plot) $\chi(t)$ (shown in blue), whose leading-order terms scale with the signal's second and third moments, respectively. (b)-(c) Survival probability $P[\Phi_{0}]$ for the mousetrap sensor as a function of the signal amplitude $\epsilon$ and frequency parameter $k$ for a zero-mean, oscillatory signal with Gaussian envelope, $b(t) = \epsilon \exp(-t^{2}/s) \sin(2\pi k t)$, with $s = 20$, $t_0 = -15$, and $t = 15$. (d)-(e) Same for the skew sensor, using absolute value of the signal $b(t) = \epsilon~\omega (t)\exp(-t^{2}/s) \sin(2\pi k t) \left[ 1 - \text{erf}(\kappa t) \right]$, with $s = 15$ and $\kappa=0.5$. Numerical (b)-(d) and adiabatic (c)-(e) calculations are shown. The black dashed line marks the threshold contour at $P[\Phi_{0}] = 0.99$.}
     \label{fig:Thresholded_sensors}
 \end{figure*}

As an example, Figure \ref{fig:Thresholded_sensors}
 illustrates two types of thresholded sensor  obtained by choosing different pairs of instantaneous eigenstates. 
 When the initial state has support on the pair of eigenvalues $\lambda_4$ and $\lambda_6$ (top row), and for zero mean signals, the accumulated phase
\begin{eqnarray}
  \chi_{4/6}(t) & = & \frac{1}{2}\int_{t_0}^t\left(\sqrt{3} + 2b - \sqrt{3-4b+4b^2}\right) d\tau \nonumber\\
  & \approx & \int_{t_{0}}^{t}\left(\left(1 + \frac{1}{\sqrt{3}}\right) b - \frac{2 b^2}{3 \sqrt{3}} + O(b^3) \right)d\tau \; ,
\end{eqnarray}
contains leading-order contributions proportional to the signal's second moment, giving rise to the {\em mousetrap} sensor. In contrast, when the initial state has support on $\lambda_6$ and $\lambda_8$ (bottom row), the accumulated phase yields
\begin{eqnarray}
  \chi_{6/8}(t) & = & \frac{1}{2} \int_{t_0}^t \left(2b-\sqrt{3-4b+4b^2} + \sqrt{3+4b+4b^2} \right)d\tau\nonumber\\
  &\approx& \int_{t_0}^t \left(\left(1+\frac{2}{\sqrt{3}}\right)b -\frac{8b^3}{9\sqrt{3}}+O(b^5)\right)d\tau \; ,
\end{eqnarray}
whose leading-order contribution is proportional to the third moment of the signal, corresponding to the {\em skew} sensor (see Fig \ref{fig:Thresholded_sensors}(a)). Here, we suppressed explicit $\tau$-dependence of $b$ for simplicity.

To further characterize the response, we compute the survival probability $P[\Phi_{0}]$ as a function of the signal amplitude $\epsilon$ and the frequency parameter $k$ for an oscillatory signal with Gaussian envelope. For the mousetrap sensor (Figs. \ref{fig:Thresholded_sensors}(b)-\ref{fig:Thresholded_sensors}(c)), we use a zero-mean, zero-skew signal $b(t) = \epsilon \exp(-t^{2}/s) \sin(2\pi k t)$, whereas for the skew sensor  (Figs. \ref{fig:Thresholded_sensors}(d)-\ref{fig:Thresholded_sensors}(e)), we 
use a zero-mean, finite skew signal of the form $b(t) = \epsilon~\omega (t)\exp(-t^{2}/s) \sin(2\pi k t) \left[ 1 - \text{erf}(\kappa t) \right]$, with $\omega(t)$ a smooth window enforcing $b(t
_{0}) = b(t_{f}) = 0$, and the $\text{erf}(t)$ factor inducing a controlled temporal skewness. The constant offset is removed so that $\int_{t_{0}}^{t_{f}} b(t) = 0$. Numerical and adiabatic results show qualitative agreement in both cases across $(\epsilon, k)$.
The thresholded rectification mechanism is not unique to the Kitaev trimer, but arises generically in coherent few-qubit sensors from spectral structure. The essential requirement is that the sensing Hamiltonian exhibits (i) an operating region in which the relevant energy-gap differences respond approximately linearly and symmetrically to the signal amplitude near $b=0$ (suppressing phase accumulation for zero-mean fields), together with (ii) a crossover to nonlinear spectral dependence beyond a threshold, generating leading sensitivity to higher-order moments of the signal above threshold. Table~\ref{tab:sensors} illustrates this mechanism across multiple one-, two-, and three-qubit Hamiltonians. For each Hamiltonian we choose an initial superposition supported on eigenstates that probe both regimes and evaluate the accumulated phase $\chi(t)$ under the same Ramsey-type protocol used in the main text. The Kitaev trimer is selected as a minimal frustrated model that additionally provides approximate omnidirectionality, which enhances robustness to variations in the direction of the incoming signal.
 
\begin{center}
    \begin{table*}[t]
    \centering
    {\renewcommand{\arraystretch}{1.4}
\begin{tabular}{||c||c||c||}
\hline
    Hamiltonian & Eigenenergies & \quad $\Psi_{0}^{\dagger} \vert \phi_0 \rangle $ \quad\\ \hline
     \multirow{2}{*}{$H = ~b(t) X$} & $\lambda_{1} = - ~b(t) $ & \multirow{2}{*}{ $ \vert + \rangle $ } \\
     & $\lambda_{2} =  ~b(t)$ & \\ \hline
     \multirow{2}{*}{$H = Z +  ~b(t) X$} & \quad $\lambda_{1} = - \sqrt{1 + b^{2}(t)}$ \quad &  \multirow{2}{*}{ $ \vert + \rangle $ } \\
      & $\lambda_{2} = \sqrt{1 +  b^{2}(t)}$ &  \\ \hline
     \multirow{4}{*}{\quad $H = ZZ +  ~b(t) (XI + IX)$ \quad} & $\lambda_1=-1$ &  \multirow{4}{*}{ $ \vert 01 \rangle + \vert 10 \rangle $ }\\
      & $\lambda_2 = 1$ & \\
      & $\lambda_3 = -\sqrt{1 + 4 b^{2}(t)}$ & \\
      & $\lambda_4 = \sqrt{1 + 4b^{2}(t)}$ & \\ \hline
      \multirow{8}{*}{ \quad $H =  X_1X_2 + Y_2Y_3 + Z_1Z_3 + \vec{b}(t) \cdot \vec{\sigma} $ \quad} & $\lambda_1 = -\sqrt{3}- b(t)$ & \multirow{8}{*}{ \quad $ \vert 100 \rangle + \vert 110 \rangle $ \quad} \\
      & $\lambda_2 = \sqrt{3}- b(t)$ & \\
      & $\lambda_3 = -\sqrt{3}+ b(t)$ & \\
      & $\lambda_4 = \sqrt{3} + b(t)$ & \\
      & \qquad$\lambda_5 = -b(t) - \sqrt{3 - 4b(t) + 4b^{2}(t)}$ \qquad & \\
      & $\lambda_6 = - b(t) + \sqrt{3 - 4b(t) + 4b^{2}(t)}$ & \\
      & $\lambda_7 = b(t) - \sqrt{3 + 4b(t) + 4b^{2}(t)}$ & \\
      & $\lambda_8 = b(t) + \sqrt{3 + 4b(t) + 4b^{2}(t)}$ & \\ \hline
\end{tabular}
\caption{Few-spin Hamiltonians, their eigenvalues and choices of initial states for quantum sensing use cases. First row: Standard sensor; Second row: Landau-Zener sensor; Third row: Dimer sensor with classical input; Fourth row: 
Kitaev trimer with classical input. The quantum mousetrap is the Ramsey interferometer constructed via an input state, $|+\rangle =  (|\lambda_4\rangle + |\lambda_6\rangle)/\sqrt{2}$.}\label{tab:sensors}}
\end{table*}
\end{center}

\begin{figure*}[t]
    \centering
    \includegraphics[scale=1]{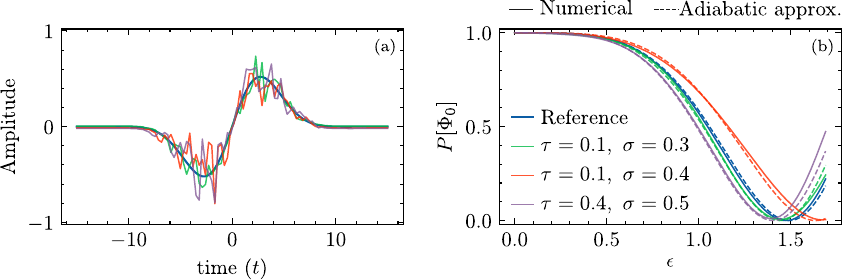}
    \caption{{\bf Robustness of the thresholded sensor under stochastic excitation.}
(a) Realization of a stochastic, zero-mean signal that follows an Ornstein–Uhlenbeck process with correlation time $\tau$, and noise amplitude, $\sigma$. (b) Corresponding survival probability $P[\Phi_0]$ for the thresholded sensor across a range of $(\tau,\sigma)$ values.}
    \label{fig:stochastic}
\end{figure*}

To test the robustness of the thresholded response, we further consider a stochastic, zero-mean Ornstein–Uhlenbeck (OU) process superimposed on the oscillatory signal, introducing temporally correlated fluctuations with adjustable correlation time, $\tau$, and amplitude, $\sigma$ (see Fig.~\ref{fig:stochastic}(a)). The corresponding survival probabilities $P[\Phi_0]$, shown in Fig.~\ref{fig:stochastic}(b), remain in close agreement with adiabatic predictions across a broad range of parameters, confirming that the sensor continues to operate effectively under zero-mean stochastic excitation. The response is bounded by the adiabatic bandwidth: slow, correlated fluctuations are averaged out, while fast noise components exceeding this bandwidth cannot be tracked adiabatically. This establishes an intrinsic operational range within which the thresholded sensor remains robust beyond idealized periodic signals.

\section{Experimental implementation}\label{app:Experimental}

\begin{figure*}
    \centering
    \includegraphics[scale=1]{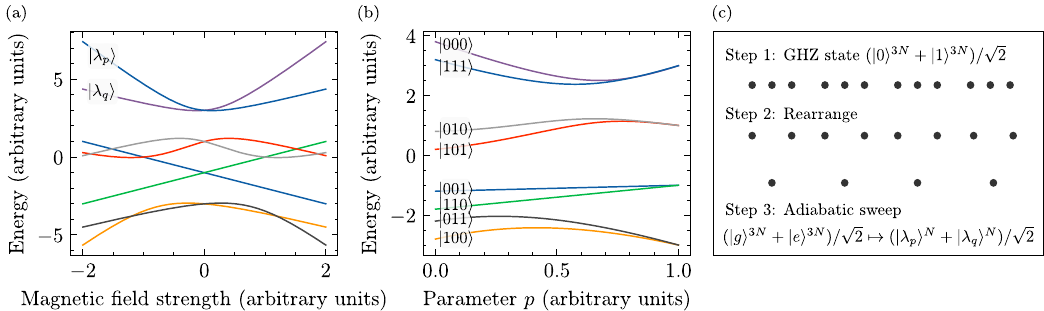}
    \caption{{\bf Potential experimental implementation.} 
    (a) Level diagram of the experimentally relevant Hamiltonian. 
    (b) Eigenenergies of $H(p)$ as a function of the interpolation parameter $p$. 
    Starting from a computational state, adiabatically sweeping $p$ from $0$ to $1$ maps it onto a desired eigenstate of the experimental Hamiltonian. 
    (c) Proposed sequence for preparing a GHZ state of trimers, where $\ket{\alpha}$ and $\ket{\beta}$ denote the target eigenstates of the trimer.}\label{expt_fig}
\end{figure*}
The geometry of the three qubits and the frustrated Heisenberg–Kitaev trimer Hamiltonian are well suited for implementation with Rydberg atoms trapped in optical tweezers. While the $ZZ$ term in the Hamiltonian can be realized via Van der Waals interaction between two Rydberg atoms, the $XX$ and $YY$ are non-trivial. Floquet engineering has been used to generate such terms in time-dependent Hamiltonians~\cite{floquetengineeringinteractionsentanglement, Zhao2023Floquet, tian2025engineeringfrustratedrydbergspin}; however, periodic driving is incompatible with the adiabatic dynamics considered here. Instead, we propose combining Fr\"{o}ster resonance~\cite{PRXQuantum.3.020303} with Van der Waals interactions.

We arrange the  atoms in an isosceles triangle with, $d_{13} \ll d_{23}=d_{12}$. A strong van der Waals coupling is engineered between atoms $1\;\&\;3$, while Fr\"{o}ster interactions are engineered between $1\;\&\;2$ and $2\;\&\;3$. Atoms 1 and 3 employ two distinct Rydberg manifolds, one addressing the van de Waals interaction and one participating in the Fr\"{o}ster interactions (See Fig.~\ref{expt_fig}). By flipping the spin in atom $1$, one maps $X_1, Y_1\mapsto X_1, -Y_1$ and $Z_1\mapsto -Z_1$, yielding the effective Hamiltonian
\begin{equation}\label{Hamiltoniain_expt}
    H_{\text{exp}}=  c_1(X_1X_2-Y_1Y_2)+c_2(X_2X_3+Y_2Y_3) + c_3Z_1Z_3 + \mathbf{b}(t)\cdot \mathbf{\sigma},
\end{equation}
which retains the spectral features required for thresholded sensing (See Fig.~\ref{expt_fig}(a)). The parameters $c_1, c_2$, and $c_3$ can be tuned using the three relative distances, the Rydberg and the two Fr\"{o}ster coupling strengths.

Scaling up the system to $N$ trimers has an additional challenge: preparing the system in the appropriate entangled state (e.g., GHZ-type state). If $\ket{\lambda_p}$ and $\ket{\lambda_q}$ are the relevant eigenstates of Eq.~(\ref{Hamiltoniain_expt}), the $N$-trimer state must be prepared as $\vert \Phi_+ \rangle =( \ket{\lambda_p}^N + \ket{\lambda_q}^N)/\sqrt{2}$. This state can be prepared adiabatically starting from a regular GHZ state. 
We introduce a tunable parameter, $p$, and define
\begin{equation}
    H(p) = (1-p)H_{\text{init}} + pH_{\text{exp}},
\end{equation}
with
\begin{equation}\label{eq:H_int}
    H_{\text{init}} = a_1 Z_1Z_2 + a_2 Z_2Z_3 + a_3 Z_1Z_3 + a_4 Z_1,
\end{equation}
where the coefficients $a_i$ are adjustable control parameters. Figure~\ref{expt_fig}(b) shows the eigenenergies of $H(p)$ as a function of $p$ for representative values $a_1 = 1$, $a_2 = 0.5$, $a_3 = 2$, and $a_4 = 0.3$.  
The eigenstates of $H_{\text{init}}$ are computational-basis states, which can be readily prepared. As $p$ is adiabatically tuned from $0$ to $1$, these states evolve continuously into the eigenstates $\ket{\lambda_p}$ and $\ket{\lambda_q}$ of $H_{\text{exp}}$.  
The eigenstates of $H_\mathrm{init}$ are the computational-basis states that can be readily prepared. Under the adiabatic interpolation of $p$ from $0$ to $1$, these eigenstates evolve continuously into the eigenstates $\ket{\lambda_p}$ and $\ket{\lambda_q}$ of $H_{\text{exp}}$, effectively mapping $\ket{000}\!\to\!\ket{\lambda_p}$ and $\ket{111}\!\to\!\ket{\lambda_q}$).  
The preparation protocol proceeds as follows:

\begin{itemize}
    \item[i.] Initialize a linear array of $3N$ qubits and prepare them in the standard GHZ state,  $\vert GHZ \rangle = (\ket{0}^{3N} + \ket{1}^{3N})/\sqrt{2}$. Such states have been demonstrated experimentally for up to $N=6$ trimers with current Rydberg-atom techniques~\cite{Omran_2019}.
    \item[ii.] Rearrange the $3N$ qubits into trimers. 
    \item[iii.] Quench the system to $H(p=0)=H_{\text{init}}$ and adiabatically sweep the parameter $p$ to $1$. 
\end{itemize}
This procedure prepares the desired GHZ-like state. After the phase acquisition stage, however, the operation must be reversed to enable readout in the computational basis. Specifically, following phase accumulation, the state takes the form $\left(\ket{\lambda_p}^N + e^{-iN\varphi}\ket{\lambda_q}^N \right)/\sqrt{2}$ where $\varphi$ denotes the acquired phase. Reversing the adiabatic sweep in step iii. maps this state to $\left(\ket{0}^{3N} + e^{-iN\varphi}\ket{1}^{3N}\right)/\sqrt{2}$. Finally, reversing the GHZ-state preparation sequence allows the phase to be read out in the computational basis, as in standard GHZ-based quantum metrology protocols.

\end{document}